\documentclass[a4paper]{article}

\usepackage{INTERSPEECH2020}
\usepackage{glossaries}
\usepackage{xcolor}
\usepackage{amsmath}
\usepackage{multirow}
\setacronymstyle{long-short}

\newacronym{ASR}{ASR}{Automatic Speech Recognition}
\newacronym{RNN-T}{RNN-T}{Recurrent Neural Network Transducer}
\newacronym{CTC}{CTC}{Connectionist Temporal Classification}
\newacronym{LAS}{LAS}{Listen-Attend-Spell}
\newacronym{BPE}{BPE}{Byte Pair Encoding}
\newacronym{WER}{WER}{Word Error Rate}
\newacronym{OOV}{OOV}{out-of-vocabulary}
\newacronym{WERR}{WERR}{Relative Word Error Rate}
\newacronym{EM}{EM}{Expectation-Maximization}
\newacronym{LM}{LM}{Language model}

\newcommand{\firstbestseg}{\sigma_{\mathrm{1best}}}

\newcommand{\x}{\mathbf{x}}
\newcommand{\y}{\mathbf{y}}
\newcommand{\w}{\mathbf{w}}

\title{Subword Regularization: An Analysis of Scalability and Generalization for End-to-End Automatic Speech Recognition}
\name{Egor Lakomkin, Jahn Heymann, Ilya Sklyar, Simon Wiesler}
\address{
  Amazon}
\email{\{egorlako, jahheyma, ilsklyar, wiesler\}@amazon.de}

\begin{document}

\maketitle
\begin{abstract}
  Subwords are the most widely used output units in end-to-end speech recognition. They combine the best of two worlds by modeling the majority of frequent words directly and at the same time allow open vocabulary speech recognition by backing off to shorter units or characters to construct words unseen during training. However, mapping text to subwords is ambiguous and often multiple segmentation variants are possible. Yet, many systems are trained using only the most likely segmentation. Recent research suggests that sampling subword segmentations during training acts as a regularizer for neural machine translation and speech recognition models, leading to performance improvements. In this work, we conduct a principled investigation on the regularizing effect of the subword segmentation sampling method for a streaming end-to-end speech recognition task. In particular, we evaluate the subword regularization contribution depending on the size of the training dataset. Our results suggest that subword regularization provides a consistent improvement of  \(2\--8\%\) relative word-error-rate reduction, even in a large-scale setting with datasets up to a size of 20k hours. Further, we analyze the effect of subword regularization on recognition of unseen words and its implications on beam diversity.
\end{abstract}
\noindent\textbf{Index Terms}: end-to-end speech recognition, regularization,  subword units

\section{Introduction}

%\jh{General remarks:
%\begin{itemize}
%\end{itemize}
%}

%\sw{open vocabulary vs large vocabulary}. <- I think its fixed now

 \par Open vocabulary \gls{ASR} systems with daily user interactions are constantly challenged by new words unseen during training.
 %Large vocabulary problem is a well-known challenge for the \gls{ASR} systems as new words are used constantly in daily interactions.
 This issue is especially pronounced for end-to-end neural models which only have a loose decomposition of acoustic and language modeling components opposed to the hybrid \gls{ASR} approach. It also makes it hard to model whole words directly since 1) the English vocabulary, for example, can exceed hundreds of thousands of words, which would dramatically increase the network size \cite{soltau17}, 2) the training data is sparse and word frequencies follow Zipfian distribution, making it difficult to get enough training samples covering each word, 3) it does not allow for \gls{OOV} words like named entities. Consequently, the first examples of large vocabulary end-to-end \gls{ASR} systems used characters as output units which in principle allow to emit any word \cite{chan2016listen, amodei2016deep} but faced challenges modeling long-term word dependencies. Subword units or interchangeably wordpieces are a middle ground between words and characters. They are currently widely used to model output units for neural machine translation \cite{sennrich2015neural} and ASR improving performance over models using character output units \cite{rao2017exploring}. 
 %where previous research suggests that a larger number of subwords in the output vocabulary improves the system's performance \cite{rao2017exploring, das2019advancing}.
 %(TODO) add a note on here about "Should mention that WP are beneficial, because has closer link to pronunciation than graphemes"
 
  \begin{figure}[t]
  \centering
  \includegraphics[width=\linewidth]{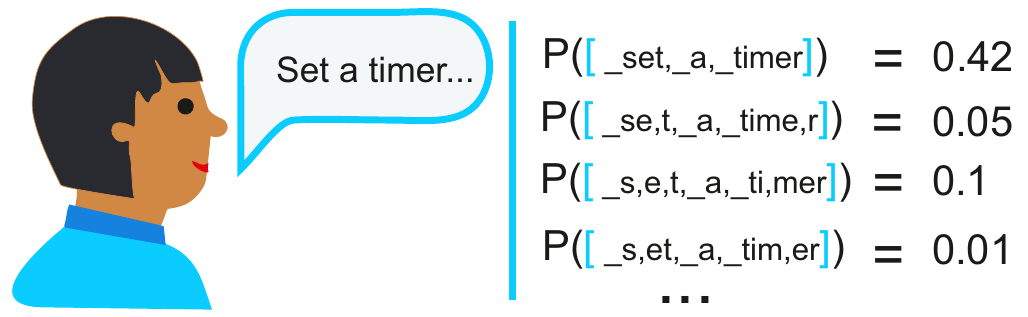}
  \caption{A transcription can be segmented into wordpieces in multiple ways. In this paper, we investigate how sampling segmentation on-the-fly can improve the accuracy of speech recognition systems.}
  \label{fig:segmentation-example}
  \vspace{-15pt}
\end{figure}
 
 Two popular methods inferring wordpiece models: \gls{BPE} \cite{schuster2012japanese, chitnis2015variable, sennrich2015neural} and a unigram wordpiece language model \cite{kudo2018subword}. The \gls{BPE} algorithm starts with only characters and iteratively merges the two most frequent symbols in the current vocabulary appearing in the training text corpora. The algorithm stops when the desired number of merge operations is reached. Kudo~\cite{kudo2018subword} introduced an alternative approach for inferring subword units by training a unigram wordpiece language model over some initial vocabulary and iteratively filtering out subwords by leaving only the ones that contribute the most to the data likelihood.% using an \gls{EM} algorithms. %TODO (maybe add the subwords + LM as an output)
 
The mapping from text to a wordpiece segmentation is not unique, cf. Fig.~\ref{fig:segmentation-example}. However, in most works, models are trained on only a single segmentation. Recent research suggests that training can be improved by varying the wordpiece segmentation during training. Kudo~\cite{kudo2018subword} introduced subword regularization: an approach for sampling wordpiece segmentations using a unigram language model. The approach has been applied to neural machine translation and lead to improvements in translation metrics across several datasets and language pairs. In another work on neural machine translation, Provilkov et al.~\cite{provilkov2019bpe} proposed \textit{BPE-Dropout}, a method  for sampling wordpiece segmentations from a \gls{BPE} model. Similar to subword regularization, \textit{BPE-Dropout} has been shown to improve the BLEU score by 3 points relative to the unregularized model and 0.9 BLUE points compared to the subword regularization.
 
In the context of \gls{ASR}, subword regularization has only been employed in a few recent publications. Hannun et al.~\cite{hannun2019sequence} obtained a \(4\%\) relative \gls{WER} reduction on LibriSpeech by sampling segmentations on word-level with a \gls{LAS} model.
 Drexler et al.~\cite{drexler19} investigated subword regularization on two end-to-end \gls{ASR} models: \gls{CTC} \cite{graves2006connectionist} and \gls{LAS}. Their results presented a \(3-7\%\) relative \gls{WER} improvement on the small-scale WSJ dataset (50 hours) but experiments with subword regularization on the larger LibriSpeech dataset (thousand hours) did not yield improvements.

 To the best of our knowledge, all the published experimental results evaluating subword regularization on speech recognition tasks were obtained on relatively small data and it remains unknown if this technique scales to larger datasets. At the same time, there is also a lack of analysis on the effect of subword regularization on the \gls{ASR} model, especially how it affects the recognition performance of unseen words during inference but also how it degrades the diversity during beam-search as different beams might just be different segmentations of the same hypothesis.
 
 %\jh{\ Maybe also mention that there is no further analysis on the effect of the regularization (e.g. OOV words, beam n-best diversity)}

 In this work, we attempt to fill this gap and present a study on how subword regularization impacts the accuracy of a streaming end-to-end \gls{ASR} system. We probe how sensitive is the method towards the output vocabulary size (e.g. total number of wordpieces). Also, we formulate a hypothesis that exploring multiple segmentations during training is beneficial for unseen words modeling. To verify our hypotheses, we conduct multiple experiments on datasets with different sizes and types: close-talk and far-field speech. Our results suggest that subword regularization method is effective for dataset sizes up-to twenty thousand hours and that the method improves the ability of end-to-end models to emit words unseen in training. 

 Our contribution is two-fold: 1) we conduct experiments evaluating the scalability of subword regularization given datasets of different size 2) we analyze the changes of the model's behavior when trained with subword regularization: namely beam decoding diversity, oracle word error rate and the ability to hypothesize unseen words.

\begin{figure}[t]
  \centering
  \includegraphics[width=\linewidth]{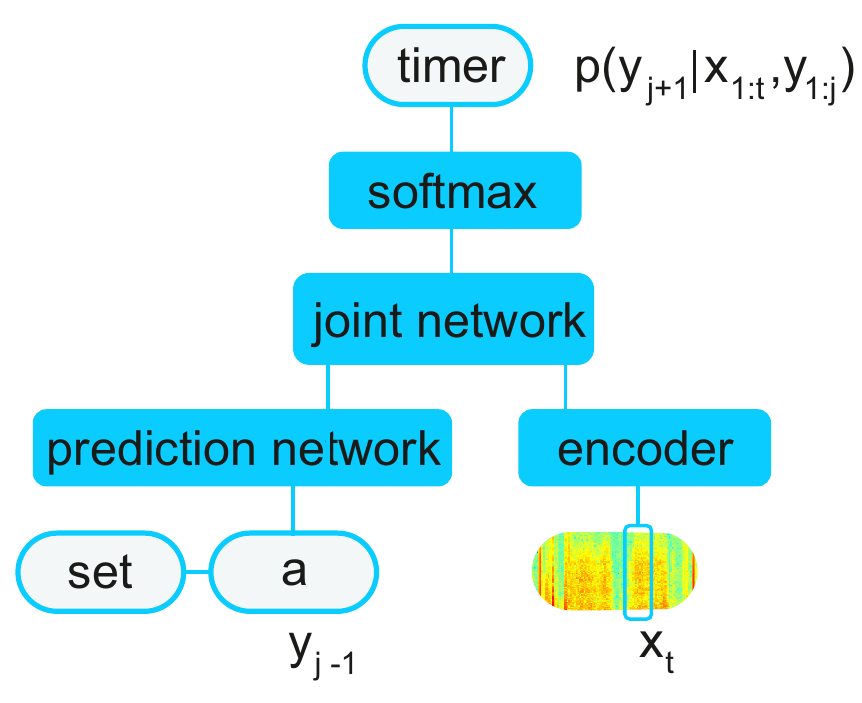}
  \caption{A schematic representation of the \gls{RNN-T} architecture and its main components. \(y_{j-1}\) is an wordpiece at the j-1 output sequence position, \(x_{t}\) is acoustic feature at the timestep $t$, and \(p(y_{j+1}|x_{1:t}, y_{1:j})\) is a probability distribution over the next output wordpiece.}
  \label{fig:speech_production}
\end{figure}

\section{Model}

% note: changed to use bold font for sequences only
In this work, we employ the \gls{RNN-T} architecture~\cite{graves2012sequence, rao2017exploring}, which is a popular model for end-to-end streaming \gls{ASR} ~\cite{he2019streaming, li2019improving, yeh2019transformer}. The \gls{RNN-T} model defines the conditional probability distribution of an output label sequence \(\mathbf{y} = (y_1, \ldots, y_U)\) given a sequence of feature vectors \(\mathbf{x} = (x_1, \ldots, x_T)\). 

The model consists of three parts: an encoder, a prediction network, and a joint network. The encoder maps the feature vectors \((x_1, \ldots, x_t)\) observed until timestep \(t\) to high-level feature representation, analogous to the acoustic model in hybrid \gls{ASR} approach:
\begin{equation}
    h^{\mathrm{enc}}_t = \mathrm{LSTM}(x_t)\;.
\end{equation}

The prediction network acts like a language model: it gets as an input an output label sequence \((y_1, \ldots, y_j)\) and outputs a fixed-length vector representations encoding information about the next possible output (see Eq. ~\ref{eq:decoder_representation}):
\begin{equation} 
\label{eq:decoder_representation}
h^{\text{dec}}_j =
\begin{cases}
 \mathrm{LSTM}(y_{j-1}) & (i) \\ 
 \mathrm{LSTM}(\hat{y}_{j-1}) & (ii) \\
\end{cases},
\end{equation}
where \((i)\) is used during training and conditions on the ground truth previous output label (teacher-forcing) and \((ii)\) is used during inference where we use the hypothesized previous output label \(\hat{y}_{j-1}\).

The joint network is a function combining transcription and prediction network representations. The common choice is a single layer feedforward network followed by a non-linearity:

\begin{equation}
    \mathcal{F}_{\text{joint}} \left(h^{\text{enc}}_t, h^{\text{dec}}_j\right) =
     \label{eq:joint}
      \mathrm{tanh}(W h^{\text{enc}}_t+V h^{\text{dec}}_j + b) \;.
\end{equation}

Finally, the conditional distribution of a next output label given the observed feature vectors and output labels is obtained by a projection to the vocabulary size and the application of softmax: 
\begin{equation}\label{eq:rnnt_prob}
\begin{split}
    \text{p} ( y_{j+1} &| x_1, \ldots, x_t, y_1, \ldots, y_j) = \\&\mathrm{softmax}\big(W'\, \mathcal{F}_{\text{joint}} \left( h^{\text{enc}}_t, h^{\text{dec}}_j\right)+b'\big)\;.
\end{split}
\end{equation}

RNN-T is typically trained to maximize the log-likelihood of the training data. In ASR, a training example is a pair of an acoustic feature sequence  \(\mathbf{x} = (x_1, \ldots, x_T)\) and a corresponding word sequence  \(\mathbf{w} = (w_1, \ldots, w_N)\). In this work, we assume that the output labels of the RNN-T system are wordpieces. Usually, only a single wordpiece segmentation corresponding to the word sequence is considered in training, e.g. the longest match for the BPE model or the most likely one with a unigram wordpiece model. We denote this mapping of a word sequence to its first best subword segmentation as \(\firstbestseg\). The training objective function is then
\begin{equation}
  L(\theta) = \mathop{\mathbb{E}_{\substack{(\mathbf{x}, \mathbf{w}) \sim P_{data}\\\mathbf{y} = \firstbestseg(\mathbf{w})}} \log P(\mathbf{y}|\mathbf{x};\theta)}\;.
  \label{sampling}
\end{equation}
Computation of the log-likelihood requires marginalizing over all possible alignments, which can be carried out by introducing a special \textit{blank} output symbol and efficiently using the forward-backward algorithm~\cite{graves2012sequence}. Since the objective function is composed of differentiable functions, the \gls{ASR} system can be trained end-to-end maximizing $L$.

% This can be calculated efficiently with the forward-backward algorithm and marginalizing over all possible alignments \cite{graves2012sequence}. Since all distributions (Eq.~\ref{eq:rnnt_prob}) are composed of differentiable functions the \gls{ASR} system can be trained end-to-end by minimizing
%\sw{mention blank label somewhere?} <- added a short note when alignment is mentioned

\section{Subword Regularization}

The idea of subword regularization is to define a probability distribution $P(\mathbf{y}|\mathbf{w})$ over the wordpiece segmentations corresponding to a word sequence $\w$. In the objective function with subword regularization, the conditional log-likelihood of a training example is replaced by an expectation w.r.t. the distribution over all segmentations:
\begin{equation}
  L(\theta) = \mathop{\mathbb{E}_{(\x,\w) \sim P_{\mathrm{data}}} \mathop{\mathbb{E}_{y \sim P(\y|\w)}\log P(\y|\x;\theta)}}\;.
  \label{loss-expectation-segs}
\end{equation}
Following Kudo~\cite{kudo2018subword}, we define $P(\mathbf{y}|\mathbf{w})$ using a wordpiece unigram language model. Typically, the segmentations are restricted to an $N$-best list. Further, the wordpiece distribution can be smoothed or sharpened using a temperature parameter~$\alpha$:
\begin{equation}\label{segmentation-sampling-proba-dist}
 P_{\alpha, N}(\mathbf{y}|\w) = \begin{cases}
 P(\y|\w)^{\alpha}/Z(\w) &\text{, if $\y$ in $N$-best}  \\
 0 &, \text{else}
 \end{cases}\;,
\end{equation}
where $Z(\w)$ is the normalization constant.
When \(\alpha=0\), segmentations are sampled from the uniform distribution. In the case of \(\alpha=\infty\), the distribution is a delta-function peaked at the most probable segmentation. 

Exact computation of the expectation $\mathbb{E}_{y \sim P(\y|\w)}$ is intractable, because the number of valid segmentations increases exponentially with the length of the word sequence. Instead, the marginalization is included in the stochastic optimization of the objective function, i.e. for every mini-batch we draw a single segmentation $\y$ from $P(\y|\w)$.

\begin{figure}[t]
  \centering
  \includegraphics[width=\linewidth]{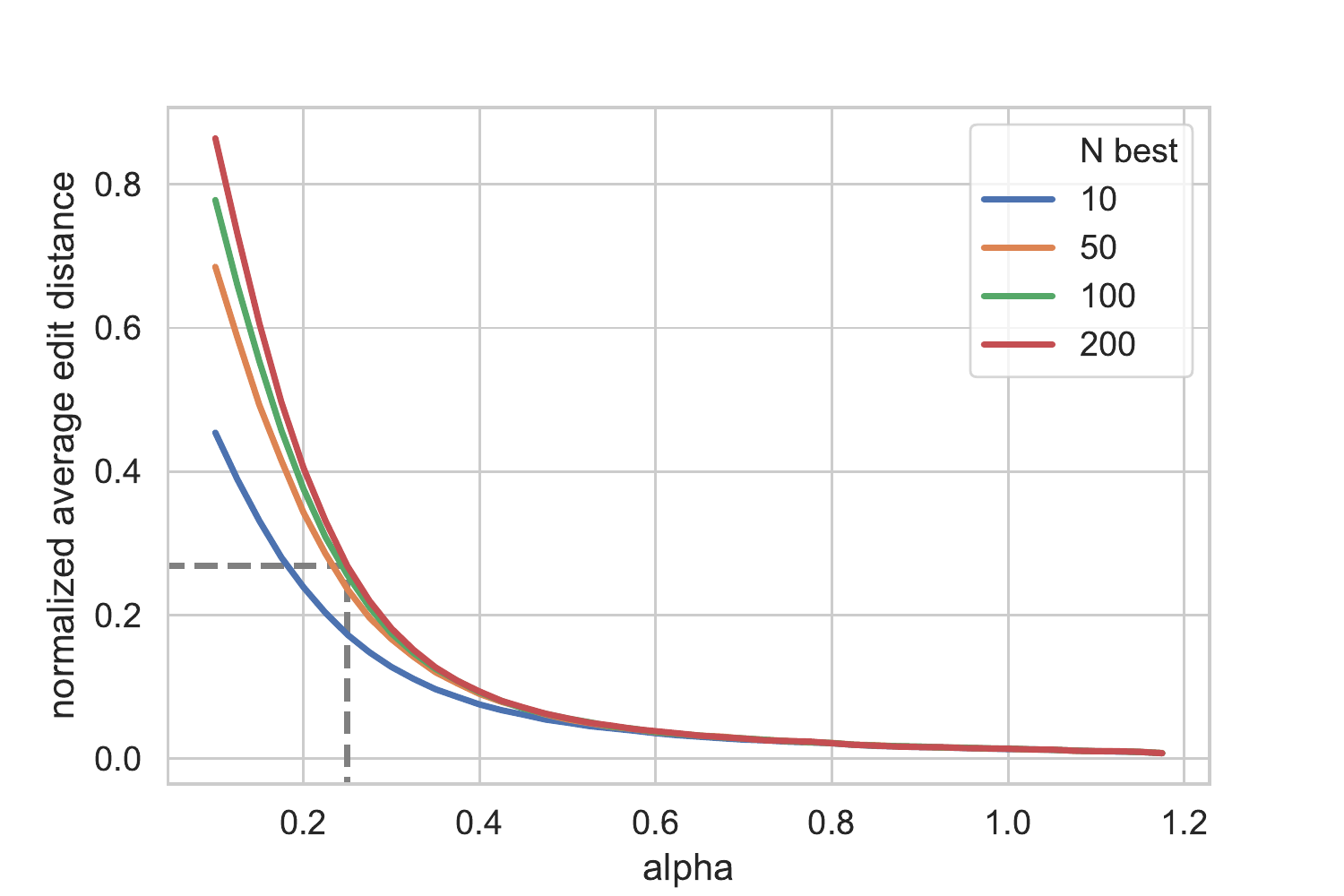}
  \caption{Expected numbers of edits per wordpiece for different number of N best segmentations depending on \(\alpha\). The black dashed lines indicates our chosen operating point which has a $26\%$ chance of an edit for a single wordpiece.}
  \label{fig:edit_chance}
\end{figure}

%\subsection{Tables}
%An example of a table is shown in Table~\ref{tab:example}. The caption text must be above the table.

%% 
%Applying (\ref{eq3}) to (\ref{eq1}), it is straightforward to see that
%% 

\section{Experiments}
%In this section we present our experimental protocol: training and testing data, model architecture and training setup parameters. We evaluate SegAugment 

\subsection{Data}

We evaluate subword regularization on two datasets: an in-house collection of real recordings of natural human interactions with voice-controlled far-field devices and the publicly available LibriSpeech corpus. Our focus is on the far-field speech recognition task.

For the far-field recognition task, we perform experiments on subsets with 2.4k, 5k, 10k, and 20k hours of transcribed audio data. Models are evaluated on two different test sets: GENERAL - a general test set following the same distribution as the training data (247k utterances), and RARE - a sample of utterances containing rare words (52k utterances).

The LibriSpeech corpus~\cite{panayotov2015librispeech} consists of 960 hours of English read speech. Models are evaluated on two test sets: test-clean and test-other.
% TODO evaluation sets

%\par Training data is sampled from Echo devices traffic for US English in 2.4k, 5k, 10k, and 20k hours.  We evaluate the models on three different test sets: Q4 - a test subset from US English Echo devices traffic (247k utterances), RW - a sample of utterances containing rate words (52k utterances), and MSG - a sample of utterances with long-form content (41k utterances). In addition, we experiment on the LibriSpeech dataset \cite{panayotov2015librispeech}, training on the 960 hours of data. We select the best model on the dev-other subset and report results on test-clean and test-other splits.

\subsection{Experimental setup}

Our \gls{RNN-T} model consists of a five-layer LSTM encoder \cite{hochreiter1997long} with 1,024 units each and a two-layer LSTM prediction network with the same number of units. We use the Adam optimizer \cite{kingma2014adam} with a warm-up, hold, and exponential learning rate decay policy and a mini-batch size of 1,536.
We extract 64-dimensional Log-Mel-Frequency features with a filterbank size of 64 every 10ms from the input speech signal. Three consecutive frames are stacked, resulting in a 192-dimensional feature vector, which is used as input to the encoder.
If not mentioned otherwise, we use an output vocabulary of 4,000 wordpieces. SpecAugment~\cite{park2019specaugment} data augmentation is performed on the unstacked features during training. In our experiments on the far-field dataset, we apply two frequency masks with a maximum size of 24 and one time mask with a maximum adaptive size of ten percent of the utterance length. In our LibriSpeech experiments, we use the LibriFullAdapt augmentation policy proposed in~\cite{park2020specaugment}.
In all our experiments with subword regularization, we used $\alpha=0.25$ and $N=200$ sampling parameters (Eq.~\ref{segmentation-sampling-proba-dist}). Figure~\ref{fig:edit_chance} illustrates the normalized average edit distance between the sampled output and the most likely segmentation with different values for \(\alpha\) and \(N\). With our settings, roughly every forth subword is changed.

All results are obtained with a pure RNN-T system without shallow fusion, language model rescoring, or context biasing as would be used in a production setup. We report results with beam search decoding with a beam width of 16.
 
\subsection{Results}

In this section, we present results on the far-field and the LibriSpeech dataset. In addition to overall results, we measure the effect of subword regularization when varying the number of wordpieces.
 
Firstly, we evaluate subword regularization on four subsets of the far-field task with 2.4k, 5k, 10k, and 20k hours of training data. The results are summarized in Table~\ref{tab:main_result}. Subword regularization consistently improves \gls{WER} for the general test set and the rare word test set. The gains are largest for the smaller datasets with up to 8\% rel. WER, but even with 20k hours we still observe a two percent relative WER improvement. Our LibriSpeech experiments are presented in Table ~\ref{tab:libri_result}. For reference, the table provides another result with \gls{RNN-T} on LibriSpeech. We observe a 2.3\% and 1.3\% relative WER improvement on \textit{test-clean} and \textit{test-other} splits.

To measure the sensitivity of subword regularization to the size of the wordpiece vocabulary, we conduct trainings with the output size of 2.5k, 4k, and 10k wordpieces. The results in Table~\ref{tab:alexa-different-wp} show that the improvement is largely independent of the vocabulary size.

%\sw{LibriSpeech results not mentioned anywhere in the text}  
% Added

\begin{table}[th]
  \vspace{-5pt}
  \caption{\gls{WERR} reductions
  with respect to the 2.4khrs model trained without subword regularization on the GENERAL and RARE test sets. We set WER to 1.0 for the baseline model performance as a reference (trained on 2.4k hours of data with 4k wordpieces). Four experiments were conducted with 2.4k, 5k, 10k, and 20k training subsets of Alexa data. Relative Word Error Reduction (\%) due to subword regularization is reported in parentheses.}
  \label{tab:main_result}
  \centering
  \begin{tabular}{ c c c c }
    \toprule
         Experiment & GENERAL & RARE \\ [0.5ex]
    \midrule
         2.4khrs & 1.00 & 2.16  \\
         +subword reg. & 0.939 (\textbf{6.1\%}) & 2.03 (\textbf{6.3\%})  \\
    \midrule   %5k Alexa exps
        5khrs & 0.893 & 1.94  \\
         +subword reg. & 0.817 (\textbf{8.43\%}) & 1.77 (\textbf{8.5\%})\\
    \midrule   %10k exps
        10khrs & 0.743 & 1.63 \\
         +subword reg. & 0.711 (\textbf{4.29\%}) & 1.57  (\textbf{3.75\%})\\
         
    \midrule  %20k exps
        20khrs & 0.653 & 1.41 \\  
         +subword reg. & 0.641 (\textbf{1.85\%}) & 1.38 (\textbf{1.85\%})\\
         
    \bottomrule
  \end{tabular}
  
\end{table}

\begin{table}[th]
  \vspace{-5pt}
  \caption{Results with RNN-T models on LibriSpeech \textit{test-clean} and \textit{test-other} splits. Numbers in parenthesis are n-best WERs.}
  \label{tab:libri_result}
  \centering
  \begin{tabular}{ c c c c c}
    \toprule
               & \multicolumn{2}{c}{WER [\%]} \\ 
    Experiment & test-clean & test-other \\ 
    \midrule
    Yeh et al. \cite{yeh2019transformer} &  12.31 & 23.16 \\
    % Zhang et al. \cite{zhang2020transformer} &  4.2 & 11.3 \\

    \midrule
         Our RNN-T & 8.16 (4.40) & 20.22 (14.67) \\
         +subword reg. & 7.97 (4.63) & 19.94 (14.81)  \\
    
    \bottomrule
    \vspace{-10pt}
  \end{tabular}
  
\end{table}

\begin{table}[th]
\vspace{-5pt}
  \caption{\gls{WERR} reductions of the models with 2.5k, 4k, and 10k wordpiece vocabulary size and subword regularization with respect to the baseline model without subword regularization (trained on 2.4khrs with 4k wordpieces).}
  \label{tab:alexa-different-wp}
  \centering
  \begin{tabular}{ c c c}
    \toprule
         Experiment & GENERAL & RARE  \\ [0.5ex]
    \midrule
         2.4khrs (4k wp) & 1.0 & 2.16 \\
         + subword reg. & 0.939  (\textbf{6.1\%}) &  2.03  (\textbf{6.3\%})\\
         \midrule
         %2.4khrs (2.5k wp) & 14.9 &  30.73 \\
         %2.4khrs (2.5k wp) + subword reg. & 14.03 &  28.97 \\
         2.4khrs (2.5k wp) & 1.05 &  2.16 \\
         + subword reg. & 0.99  (\textbf{5.8\%}) &  2.04  (\textbf{5.7\%})\\
         \midrule
         2.4khrs (10k wp). & 1.03 &  2.17 \\
         + subword reg. & 0.97  (\textbf{6.2\%}) &  2.03  (\textbf{6.3\%\%})\\

    \bottomrule
  \end{tabular}
  \vspace{-17pt}
\end{table}

\section{Analysis}

\subsection{Modeling unseen words}

Even though our subword-based \gls{ASR} system is open vocabulary, we expect a lower accuracy for words that haven't been present in the audio training data. The model hasn't observed the pronunciation of the full word and, in general, will assign a low probability for unseen combinations of subword units. This is a particularly important problem for end-to-end ASR systems in comparison to HMM-based ASR systems as pronunciations cannot be inferred from a lexicon. Further, the end-to-end system doesn't make use of text training data, which could be beneficial to infer the language model context for unseen words. Even though it is in principle possible to apply a language model using shallow fusion, its effectiveness is limited because the acoustic and language model are not fully decoupled in RNN-T.

The model trained with subword regularization is exposed to a larger variety of wordpiece sequences. Hence, we hypothesize that the \gls{ASR} model is  more capable of producing words unseen during training. We verify this by analyzing the decoding results of models trained with and without subword regularization on the rare word test set of the far-field task. This has a relatively large fraction of unseen words (1.93\%, 1.27\%, and 0.85\% overall for 5k, 10k, and 20k experiments correspondingly).

In the analysis, we compute precision, recall, and F-score metrics by treating unseen word detection as a binary classification problem. We calculate how many times the model emits a correct (true positive, $\mathrm{tp}$) or incorrect (false positive, $\mathrm{fp}$) unseen word on the test set, and the number of times the model does not emit the correct unseen word (false negative, $\mathrm{fn}$). We calculate precision \({\mathrm{tp}}/{(\mathrm{tp}+\mathrm{fp})}\), recall \({\mathrm{tp}}/{(\mathrm{tp}+\mathrm{fn})}\) and F-score \(\left(2*\frac{\mathrm{precision}*\mathrm{recall}}{\mathrm{precision}+\mathrm{recall}}\right)\) metrics to evaluate the unseen word recognition performance.
The results are shown in Table~\ref{tab:oov_results}. We observe gains in unseen word recognition with subword regularization across all metrics. The overall F-score improvement ranges from 9\% to 28\%. As expected, the largest gains are obtained on the smallest dataset.

\begin{table}[th]
  \vspace{-6pt}
  \caption{Precision, Recall and F-score of unseen word recognition with 5k, 10k, and 20k hour training datasets.}
  \label{tab:oov_results}
  \centering
  \begin{tabular}{ c c c c}
    \toprule
         Model & Precision & Recall & F-score \\ [0.5ex]
    \midrule
         5khrs & 0.06 & 0.09 & 0.07 \\
         5khrs + sampling & 0.07 & 0.12 & \textbf{0.09 (+28.5\%)} \\
    \midrule
         10khrs & 0.07 & 0.12 & 0.08 \\
         10khrs + sampling & 0.08 & 0.14 & \textbf{0.10 (+25.0\%)} \\
    \midrule
         20khrs & 0.09 & 0.15 & 0.11 \\
         20khrs + sampling & 0.09 & 0.18 & \textbf{0.12 (+9\%)} \\
    \bottomrule
  \end{tabular}
  \vspace{-12pt}
\end{table}

Our hypothesis on why the subword regularization is effective for modeling unseen words is that it shifts the label distribution such that smaller wordpiece units such characters appear more often during training. To verify this, we compute frequencies of each wordpiece emitted decoding the RARE words test set and group them by length. The only significant change we note for models with subword regularization is the increase of single character wordpieces usage frequency by 24\% and 17\% 
%\jh{is this the increase in usage? If not, what are the baseline numbers?} 
for 10k and 20k experiments. This result can be expected as even not all wordpieces with the length of two are present in the output vocabulary, which makes characters the only choice for the model to emit an unseen word correctly.

%TODO: (separate note on that could be added) reduce the mismatch between training and inference caused by the teacher-forcing.  

\subsection{N-best WER and beam diversity}

In our experiments, we observed that the n-best \gls{WER} does not improve in contrast to the first best \gls{WER} as a side effect of subword regularization. We analyzed the hypotheses in the decoding beam and discovered that there is a significant amount of duplicate hypotheses. For example, for the \textit{10khrs, 20khrs} experiments without sampling, there are on average 97.5\% and 96.8\% unique hypotheses in the 16-best beam respectively, while for the models with subword regularization the ratio of unique hypotheses drops to 56.9\% and 54.8\%. This directly translates to a degradation in the n-best \gls{WER} which degrades by 9.6\% and 14.0\% rel. \gls{WER} for \textit{10khrs} and \textit{20khrs} experiments, even though the first best \gls{WER} improves. We observe a similar change in our LibriSpeech experiments: with subword regularization, the ratio of unique hypotheses on \textit{test-clean} reduces from 94\% to 71.3\%. This behavior is expected for the default beam decoding implementation as the model with subword regularization is trained to explore alternative segmentations and therefore assigns a more balanced probability mass to segmentations corresponding to the same word sequence.

%\subsection{RNN-T loss distribution with character segmentations}

 %- sentence with the OOV word, plot loss with alignment
 %- split into characters and calculate RNN-T loss for models with and %without sampling
%\par In addition, we analyzed the distribution of \gls{RNN-T} loss

\section{Conclusions}

We presented a study on subword regularization applied to streaming end-to-end \gls{ASR}. Our results suggest that sampling wordpiece segmentations on-the-fly during training improves word error rate across several setups and dataset sizes from 1k to 20k hours. We demonstrate that subword regularization improves recognition accuracy of unseen words which is driven by the enhanced ability of an \gls{ASR} model to utilize character units.

We leave for a future work an investigation of changes to the beam decoding required to utilize the potential of subword regularization. As we observed that almost 50\% of hypotheses in the final beam are non-unique which impacts oracle \gls{WER} which limits improvements from rescoring methods. Also, we plan to investigate combining several segmentations of an utterance transcription in the loss function to get a better estimation of the probability of a word sequences given the acoustic input. 

\bibliographystyle{IEEEtran}

\bibliography{mybib}

\begin{thebibliography}{10}
\providecommand{\url}[1]{#1}
\csname url@samestyle\endcsname
\providecommand{\newblock}{\relax}
\providecommand{\bibinfo}[2]{#2}
\providecommand{\BIBentrySTDinterwordspacing}{\spaceskip=0pt\relax}
\providecommand{\BIBentryALTinterwordstretchfactor}{4}
\providecommand{\BIBentryALTinterwordspacing}{\spaceskip=\fontdimen2\font plus
\BIBentryALTinterwordstretchfactor\fontdimen3\font minus
  \fontdimen4\font\relax}
\providecommand{\BIBforeignlanguage}[2]{{%
\expandafter\ifx\csname l@#1\endcsname\relax
\typeout{** WARNING: IEEEtran.bst: No hyphenation pattern has been}%
\typeout{** loaded for the language `#1'. Using the pattern for}%
\typeout{** the default language instead.}%
\else
\language=\csname l@#1\endcsname
\fi
#2}}
\providecommand{\BIBdecl}{\relax}
\BIBdecl

\bibitem{soltau17}
H.~Soltau, H.~Liao, and H.~Sak, ``{Neural {Speech} {Recognizer}:
  Acoustic-to-Word {LSTM} Model for Large Vocabulary Speech Recognition},'' in
  \emph{Proc. Interspeech}, 2017, pp. 3707--3711.

\bibitem{chan2016listen}
W.~Chan, N.~Jaitly, Q.~Le, and O.~Vinyals, ``{Listen, {Attend} and {Spell}: A
  Neural Network for Large Vocabulary Conversational Speech Recognition},'' in
  \emph{Proc. ICASSP}.\hskip 1em plus 0.5em minus 0.4em\relax IEEE, 2016, pp.
  4960--4964.

\bibitem{amodei2016deep}
D.~Amodei, S.~Ananthanarayanan, R.~Anubhai, J.~Bai, E.~Battenberg, C.~Case,
  J.~Casper, B.~Catanzaro, Q.~Cheng, G.~Chen \emph{et~al.}, ``{Deep {Speech} 2:
  End-to-end Speech Recognition in {English} and {Mandarin}},'' in \emph{Proc.
  ICML}, 2016, pp. 173--182.

\bibitem{sennrich2015neural}
R.~Sennrich, B.~Haddow, and A.~Birch, ``{Neural Machine Translation of Rare
  Words with Subword Units},'' in \emph{Proc. ACL}, 2016.

\bibitem{rao2017exploring}
K.~Rao, H.~Sak, and R.~Prabhavalkar, ``{Exploring Architectures, Data and Units
  for Streaming End-to-End Speech Recognition with {RNN}-Transducer},'' in
  \emph{Proc. ASRU}, 2017, pp. 193--199.

\bibitem{schuster2012japanese}
M.~Schuster and K.~Nakajima, ``Japanese and {Korean} voice search,'' in
  \emph{Proc. ICASSP}.\hskip 1em plus 0.5em minus 0.4em\relax IEEE, 2012, pp.
  5149--5152.

\bibitem{chitnis2015variable}
R.~Chitnis and J.~DeNero, ``{Variable-length Word Encodings for Neural
  Translation Models},'' in \emph{Proc. EMNLP}, 2015, pp. 2088--2093.

\bibitem{kudo2018subword}
T.~Kudo, ``{Subword {Regularization}: Improving Neural Network Translation
  Models with Multiple Subword Candidates},'' in \emph{Proc. ACL}, 2018, pp.
  66--75.

\bibitem{provilkov2019bpe}
I.~Provilkov, D.~Emelianenko, and E.~Voita, ``{{BPE}-{Dropout}: Simple and
  Effective Subword Regularization},'' \emph{arXiv preprint arXiv:1910.13267},
  2019.

\bibitem{hannun2019sequence}
A.~Hannun, A.~Lee, Q.~Xu, and R.~Collobert, ``{Sequence-to-Sequence Speech
  Recognition with Time-Depth Separable Convolutions},'' in \emph{Proc.
  Interspeech}, 2019, pp. 3785--3789.

\bibitem{drexler19}
J.~Drexler and J.~R. Glass, ``{Subword Regularization and Beam Search Decoding
  for End-to-End Automatic Speech Recognition},'' in \emph{Proc. ICASSP}, 2019,
  pp. 6266--6270.

\bibitem{graves2006connectionist}
A.~Graves, S.~Fern{\'a}ndez, F.~Gomez, and J.~Schmidhuber, ``{Connectionist
  {Temporal} {Classification}: Labelling Unsegmented Sequence Data with
  Recurrent Neural Networks},'' in \emph{Proc. ICML}, 2006, pp. 369--376.

\bibitem{graves2012sequence}
A.~Graves, ``{Sequence Transduction with Recurrent Neural Networks},''
  \emph{arXiv preprint arXiv:1211.3711}, 2012.

\bibitem{he2019streaming}
Y.~He, T.~N. Sainath, R.~Prabhavalkar, I.~McGraw, R.~Alvarez, D.~Zhao,
  D.~Rybach, A.~Kannan, Y.~Wu, R.~Pang \emph{et~al.}, ``{Streaming End-to-End
  Speech Recognition for Mobile Devices},'' in \emph{Proc. ICASSP}, 2019, pp.
  6381--6385.

\bibitem{li2019improving}
J.~Li, R.~Zhao, H.~Hu, and Y.~Gong, ``{Improving {RNN} Transducer Modeling for
  End-to-End Speech Recognition},'' in \emph{Proc. ASRU}, 2019, pp. 114--121.

\bibitem{yeh2019transformer}
C.-F. Yeh, J.~Mahadeokar, K.~Kalgaonkar, Y.~Wang, D.~Le, M.~Jain, K.~Schubert,
  C.~Fuegen, and M.~L. Seltzer, ``{Transformer-{Transducer}: End-to-End Speech
  Recognition with Self-Attention},'' \emph{arXiv preprint arXiv:1910.12977},
  2019.

\bibitem{panayotov2015librispeech}
V.~Panayotov, G.~Chen, D.~Povey, and S.~Khudanpur, ``{LibriSpeech: an {ASR}
  Corpus based on Public Domain Audio Books},'' in \emph{Proc. ICASSP}.\hskip
  1em plus 0.5em minus 0.4em\relax IEEE, 2015, pp. 5206--5210.

\bibitem{hochreiter1997long}
S.~Hochreiter and J.~Schmidhuber, ``Long {Short}-{Term} {Memory},''
  \emph{Neural computation}, vol.~9, no.~8, pp. 1735--1780, 1997.

\bibitem{kingma2014adam}
D.~P. Kingma and J.~Ba, ``{Adam: {A} Method for Stochastic Optimization},'' in
  \emph{Proc. ICLR}, 2015.

\bibitem{park2019specaugment}
D.~S. Park, W.~Chan, Y.~Zhang, C.~Chiu, B.~Zoph, E.~D. Cubuk, and Q.~V. Le,
  ``{SpecAugment: {A} Simple Data Augmentation Method for Automatic Speech
  Recognition},'' in \emph{Proc. Interspeech}, 2019, pp. 2613--2617.

\bibitem{park2020specaugment}
D.~S. Park, Y.~Zhang, C.-C. Chiu, Y.~Chen, B.~Li, W.~Chan, Q.~V. Le, and Y.~Wu,
  ``{SpecAugment on Large Scale Datasets},'' \emph{ICASSP 2020 - 2020 IEEE
  International Conference on Acoustics, Speech and Signal Processing
  (ICASSP)}, May 2020.

\end{thebibliography}

% \begin{thebibliography}{9}
% \bibitem[1]{Davis80-COP}
%   S.\ B.\ Davis and P.\ Mermelstein,
%   ``Comparison of parametric representation for monosyllabic word recognition in continuously spoken sentences,''
%   \textit{IEEE Transactions on Acoustics, Speech and Signal Processing}, vol.~28, no.~4, pp.~357--366, 1980.
% \bibitem[2]{Rabiner89-ATO}
%   L.\ R.\ Rabiner,
%   ``A tutorial on hidden Markov models and selected applications in speech recognition,''
%   \textit{Proceedings of the IEEE}, vol.~77, no.~2, pp.~257-286, 1989.
% \bibitem[3]{Hastie09-TEO}
%   T.\ Hastie, R.\ Tibshirani, and J.\ Friedman,
%   \textit{The Elements of Statistical Learning -- Data Mining, Inference, and Prediction}.
%   New York: Springer, 2009.
% \bibitem[4]{YourName17-XXX}
%   F.\ Lastname1, F.\ Lastname2, and F.\ Lastname3,
%   ``Title of your INTERSPEECH 2020 publication,''
%   in \textit{Interspeech 2020 -- 20\textsuperscript{th} Annual Conference of the International Speech Communication Association, September 15-19, Graz, Austria, Proceedings, Proceedings}, 2020, pp.~100--104.
% \end{thebibliography}

\end{document}